# CHANGE OF MASS IN THE CONSERVATIVE SYSTEM OF BODIES

**A.V. Glushkov**

*Yu.G. Shafer Institute of Cosmophysical Research and Aeronomy,*
*31 Lenin Ave., 677980 Yakutsk, Russia*

e-mail: a.v.glushkov@ikfia.ysn.ru

A problem of mass in macro- and microcosm has been considered from the single point of view on the basis of the law of conservation of energy. It is shown that in the conservative (absolutely closed) system all types of motion and interaction of the bodies are possible only at the expense of the mass defect of bodies themselves. As the result, their mass becomes less than the initial rest mass before forming the system. It leads to some change in the energy balance of the system and it is noticeably manifested in the motion of bodies in the relativistic velocity region. The accounting of that essentially adds a customary picture for the motion of bodies in the conservative system.

## 1. Introduction

The mass is one of basic physical characteristics of a matter determining its inert and gravitational properties. In the classical mechanics the mass is the coefficient of proportionality between the force affecting on a body and its acceleration – in this case it is called as an inert one. Besides, the mass creates the gravitational field and it is called as the gravitational mass. The inert and gravitational masses are equal to each other (equivalence principle).

The relativity theory has slightly opened one more very important property of the mass, i.e. its equivalence to the total energy of body which is expressed in the famous Einstein formula $E = mc^2$. From this moment the energy $E$ and the mass $m$ are combined in the new fundamental concept: energy-mass. The mass is not an additive value. For example, when two initial particles form a new steady system, then the mass defect occurs − the energy $\Delta E$ is released which corresponds to the change of system mass by $\Delta m = \Delta E/c^2$. The mass defect defines a picture of microcosm's in many respects, but it exists also in macrocosm.

The mass is the manifestation of main interaction of a body with the World medium. It is accessible to our observation in a very wide range: almost from zero up to the size of the Metagalaxy (~ $10^{52}$ kg) [1]. Sometimes the role of the mass is underestimated in those or other processes, which seem to be quite clear and explainable. Some sites of manifestation of the mass in absolutely closed (conservative) system will be considered below which spill the additional light on this problem.

## 2. On the time

In our further reasoning we shall use such important concepts as space and time, which are badly defined up till now and cause different interpretation. We shall define them from the philosophical point of view: space as an ideal object is born in the process of extreme fast ("instantaneous") reflection of the objects of world on the basic object − the coordinate system. Similarly, the time is the result of reflection, the relationship of processes of motion and development to some process selected as the basic one.

When we speak about the spatial-temporary continuum (for example, in the relativity theory), it means that we describe the world, giving the relative coordinates of objects and relative velocities of processes in these objects (current of time) from the point of view of the observer in the chosen coordinate system. Therefore it is naturally to believe that the spatial and temporary coordinates are disparate, nonequivalent, and the four-dimension of space-time is of the form 3 + 1.

In the real world there is one and unique original cause – the mutual actions of the material medium. Depending on it, the time is determined. The action which occurred, occurs or will occur defines the time by the following concepts: last, present or future. The concept of the time exists only in the abstract world of a man and doesn't have the physical essence. So, a sand glass and a pendulum clock stop to go in weightlessness, but it does not mean that the time stops. Here a choice of the time standard plays the important role. If it is the atomic clock, then the duration of action equal to 9192631770 periods of radiation is accepted for 1 second that corresponds to the transition between two hyperfine levels of the ground state of the caesium 137 atom [1].

According to the relativity theory, in a body moving with a velocity $v$, the period of hours $T$ is increased in terms of the Lorenz transformation:

$$T = T_0/\sqrt{1 - \beta^2} , \qquad (1)$$

where $T_0$ is the period of the immovable control clocks; $\beta = v/c$ is the dimensionless velocity of motion of a body in space ($c$ is the velocity of light). The validity of (1) has been proved by the numerous experiments. However the intervals of time themselves

$$\Delta t \sim 1/T \sim N, \quad \Delta t_0 \sim 1/T_0 \sim N_0 \qquad (2)$$

can be estimated only by the number $N_0$ of "revolutions of the hand" of a clock (or by the number of atomic periods), which will be required for being done of the any action in the stationary (laboratory) system of coordinates and by the number of revolutions $N$ of the precisely same clock located on a moving body. For example, under terrestrial conditions one second is equal to $N_H = 6.579\,683\,786\,153 \times 10^{15}$ periods of rotations of an electron around of the centre of unexcited hydrogen atom [2]. Under any other conditions, if the hydrogen clocks are used, we should also take $N_H$ revolutions of the electron for one second. It is obvious that the periods of both clocks $T$ and $T_0$ are associated with their indications $N$ and $N_0$ by the following equality:

$$NT = N_0 T_0 . \tag{3}$$

Let $m_0$ is the mass of the stationary control balance, and $E_0 = m_0 c^2$ is their total energy. The energy of moving clocks $E$, according to the relativity theory, will be higher:

$$E = mc^2 = E_0/\sqrt{1-\beta^2} . \tag{4}$$

Then for the relation of the periods it is possible to write:

$$T/T_0 = E/E_0, \text{ or } T \sim E, T_0 \sim E_0 . \tag{5}$$

Thus, the atomic period of the tested clocks is increased directly proportionally to the total energy of these clocks. It is obvious, it will be also correctly in other cases, when the energy increases, for example, on being heated (transfer of heat) or compressed (occurrence of mechanical stress). In general case, to change the atomic period $T_0$ and the total energy $E_0$ of a clock it is a necessary to make any work or to transfer energy to them, which can, at least, change the kinetic energy of the elementary oscillators (atomic clock).

Using (3) and (5), for the clock indications $N_0$ and $N$ we obtain:

$$N/N_0 = T_0/T = E_0/E, \text{ or } N \sim 1/E, N_0 \sim 1/E_0 , \tag{6}$$

that is the clock indications decrease as the total energy of clock increases. If we compare simultaneously some tested clock and stationary control one in the inertial system, then a ratio of their indications (6) can be called as a relative velocity of current time $\gamma$:

$$\gamma = N/N_0 = T_0/T = E_0/E . \tag{7}$$

For a case, when the tested clock moves with a constant velocity $v$ we have:

$$\gamma = \sqrt{1-\beta^2} . \tag{8}$$

As $\gamma < 1$, then the velocity of a course of time of the moving clock is less than the velocity of a course of time of the control (stationary) clock, or the time in moving clock is delayed.

The velocity of light is the fundamental value determining many physical processes. As Einstein showed [3], it depends on the gravitational potential. For the further analysis we shall introduce some designations, which we shall use below. We shall designate the velocity of light in the vacuum at a distance $R$ from the centre of the gravitating mass $M$ by the symbol "$c(R)$", and $c_0$ corresponds to the velocity of light in vacuum far from all gravitating masses. Both these velocities are connected among themselves by the relationship

$$c(R) = (c(R)/c_0)c_0 = (\Delta t_0/\Delta t(R))c_0 , \qquad (9)$$

where $\Delta t_0$ and $\Delta t(R)$ are the lengths of time in the corresponding points of space necessary for the light to pass the same way $\Delta l$ with the velocities $c = \Delta l/\Delta t$ and $c_0 = \Delta l/\Delta t_0$.

From (7) and (9) it follows, that

$$c(R)/c_0 = N_0/N(R) = \gamma_{gr}(R) , \qquad (10)$$

where $N(R)$ is the number of atomic periods fixed in a given point of gravitational field by the external observer according to clock, located in this point of the field; $N_0$ is the number of periods in this point in absence of a gravitational field (or indication of the atomic clock far from a gravitational mass $M$). According to Einstein [3,4]:

$$c(R)/c_0 = 1 - |\Phi(R)| = \gamma_{gr}(R) . \qquad (11)$$

Here $\Phi(R) = \varphi(R)/(c_0)^2$ is the dimensionless gravitational potential. The function $\varphi(R)$ is determined by the well-known Newton formula:

$$\varphi(R) = -GM/R , \qquad (12)$$

where $G = 6.672 \times 10^{-11}$ kg$^{-1}$m$^3$s$^{-2}$ is the gravitational constant. The gravitational potential is the work that must be performed over an unit mass to remove it from the gravitational field. From (11) the equation of motion in the gravitational field follows

$$1 = \gamma_{gr}(R) + |\Phi(R)| , \qquad (13)$$

where is convenient to solve the number of tasks. For example, if we multiply its both parts by $E_0 = m_0(c_0)^2$, we obtain:

$$E_0 = \gamma_{\text{gr}}(R)E_0 + |\Phi(R)|E_0 = const. \qquad (14)$$

The first term on the right is equal to the total energy of body

$$E(R) = \gamma_{\text{gr}}(R)E_0 = m(R)(c_0)^2 \qquad (15)$$

with the mass

$$m(R) = \gamma_{\text{gr}}(R)m_0 \qquad (16)$$

in the gravitational field, if we transfer it slowly from infinity to the given point of the field. The second term of the equation (14):

$$|\Phi(R)|E_0 = m_0|\varphi(R)| = U(R) \qquad (17)$$

characterizes the potential energy of the gravitational field, which is equal to:

$$U = (1 - \gamma_{\text{gr}})E_0 = \Delta m_0(c_0)^2 = \Delta E_0. \qquad (18)$$

The formulae (10) and (15) are connected among themselves by the relationship:

$$\gamma_{\text{gr}}(R) = E(R)/E_0 = N_0/N(R) \leq 1, \qquad (19)$$

from which it follows, that in the increasing gravitational field the total energy of clock $E(R)$ decreases, and the elementary oscillators increase their cyclic frequency $N(R)$, i.e. the atomic clock go faster. In particular, if the observer will be more deeply in the gravitational field, than the tested clock ($R_1 < R_2$), he will find out an opposite effect – his own atomic clock (in a point $R_1$) will go faster than the one located further ($N(R_1) > N(R_2)$).

## 3. Equation of Motion

For the further reasoning it is convenient to transform (8) into the equation:

$$1 = \gamma^2 + \beta^2, \qquad (20)$$

which reflects the motion of the end of an unit radius-vector of a velocity $\boldsymbol{j} = \beta + i\gamma$ along a circle around the origin of coordinates in the complex plane. It contains considerable and allows, for example, to solve easy some difficult tasks of the relativistic mechanics. So, if we multiply both parts of (20) by $E^2$, we obtain:

$$E^2 = (\gamma E)^2 + (\beta E)^2 . \tag{21}$$

With allowance for (4) and $mv = p$ (impulse of motion of a body in the laboratory system of coordinates) we have the well-known relativity theory equation:

$$E^2 = (E_0)^2 + (pc)^2 . \tag{22}$$

And if we multiply both parts of (20) by $(c\Delta t)^2$, then we obtain the relationship:

$$(c\Delta t)^2 = (c\Delta t\gamma)^2 + (c\Delta t\beta)^2 = (c\Delta t_0)^2 + (v\Delta t)^2 , \tag{23}$$

in which another not less known, relativity theory equation, i.e. spatial-temporal interval is easily guessed:

$$(\Delta s)^2 = (\Delta l)^2 - (\Delta r)^2 = inv . \tag{24}$$

Usually it is considered as a geometrical factor of the curved four-dimensional space-time [1]. However, it can also have some other physical sense (see section 6). In Fig. 1 the alternative geometrical interpretation of the interval (24) is given. It is seen that $\Delta s = c\Delta t_0$ is the distance, which is covered by the matter with the velocity of light $c$ in the distinctive internal motion for the time $\Delta t_0$ kept by the clock, moving together with a body with the velocity $v$, when the light and the body pass the ways $\Delta l = c\Delta t$ and $\Delta r = v\Delta t$ in the external medium for the time $\Delta t$ kept by an observer by stationary clock. The dashed line of arbitrary length in Fig. 1 explains the essence of this invariant: an invariance of $\Delta s$ at any velocity $v$.

The following rations

$$\Delta s/\Delta l = \Delta t_0/\Delta t = \gamma = \cos\theta , \qquad \Delta r/\Delta l = v/c = \beta = \sin\theta$$

characterize the certain form of the motion itself of the matter. From the equations (20) and (24) it is seen that this motion actually satisfies the Pythagor theorem, which characterizes apparently, one of the most fundamental properties of the matter structure (from micro- to macro-cosm).

Now we consider the problem of the mass of body. The answer to the question on the change of mass of a moving body has been partly given already by the relativity theory. It is in the formula (4), which is summed of the rest energy-mass of a body $E_0 = m_0c^2$ and the kinetic energy of motion $K = m_0v^2/2$. The body obtains $K$ outside from a driver accelerating the given

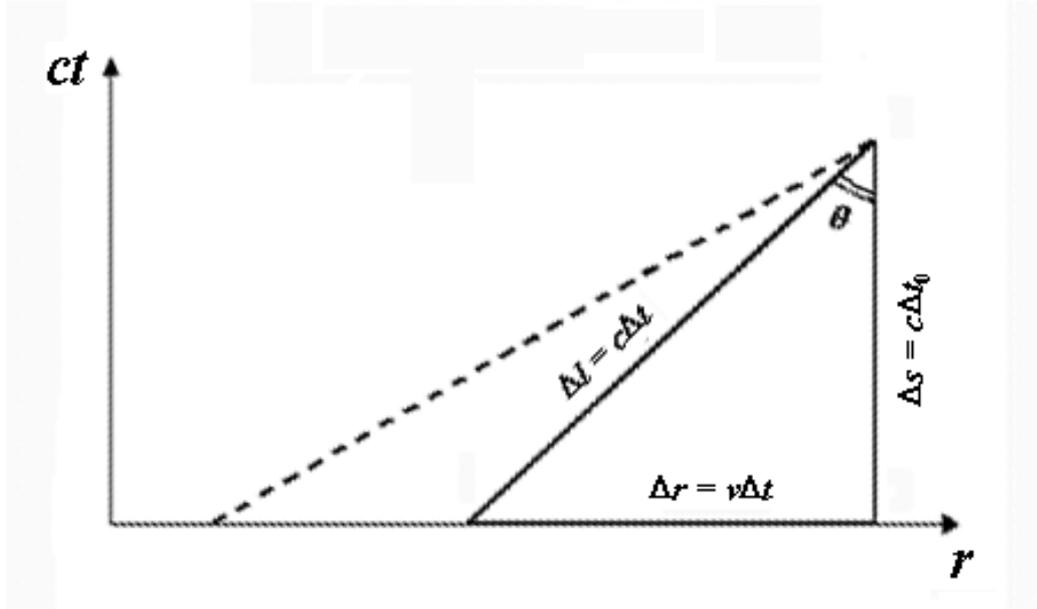

**Fig. 1.** Scheme explaining the sense of the spatial-temporal interval (24) within the framework of the Pythagor theorem; the dashed line is the hypotenuse of arbitrary length.

body to the velocity $v$ (curve 3 in Fig. 2). That is

$$E = mc^2 = E_0 + K. \qquad (25)$$

Hence it's seen that the mass

$$m = m_0 + \Delta m = m_0 + K/c^2 \qquad (26)$$

of the moving body $m > m_0$. The similar conclusion directly follows from the equation (4), if we divide its both parts by $c^2$:

$$m = m_0/\sqrt{1 - \beta^2} = m_0/\gamma. \qquad (27)$$

However, the increasing value (27) characterizes not all changes of the mass of body. So, the formula (16) shows that in the gravitational field the mass can decrease. There are also other systems, where the initial rest mass of the body decreases. Consider some of them. In order to understand better the essence of the problem in the change of a body mass, we use again the motion equation (20). Let's multiply its both parts by $E$:

$$E = \gamma^2 E + \beta^2 E = \gamma E_0 + E_0 \beta^2/\gamma = E_T + E_R. \qquad (28)$$

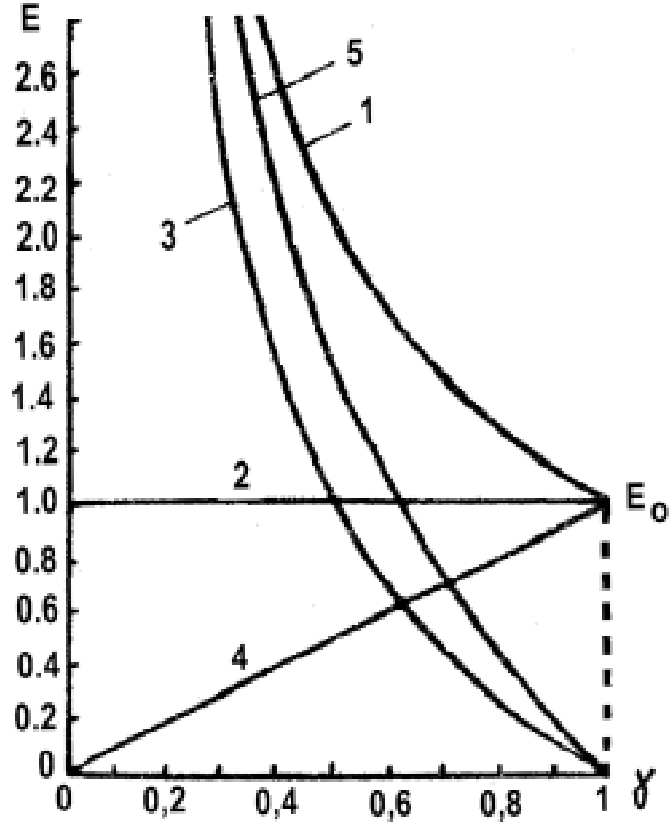

**Fig. 2.** Motion energy of a body of different kinds of the Lorenz factor $\gamma$. *1* is $E = E_0/\gamma = E_0 + K = E_T + E_R$ (total energy); *2* is $E_0 = m_0 c^2$ (rest mass energy); *3* is $K = E - E_0$ (kinetic energy); *4* is $E_T = \gamma E_0 = \gamma^2 E$ (internal motion energy of a body in time); *5* is $E_R = E_0 \beta^2/\gamma = E_\Delta + K$ (motion energy of a body in space).

As a result, we obtain the new equation different from (25). It reveals the character of the energy changes occurring in a moving body more fully. Let as pay attention at the component $E_T$:

$$E_T = \gamma^2 E = \gamma E_0 = (\gamma m_0)c^2 = mc^2. \tag{29}$$

Let's name it as the energy of internal motion of a body in time. The plot of dependence $E_T$ on $\gamma$ is shown in Fig. 2 by the strait inclined line *4*. It is seen that this energy in a moving body is always less than the rest energy $E_0$ of a body. In this case, the faster a body goes, the less $E_T$ is. And it occurs, according to (29), at the expense of reduction of the primary mass $m_0$ of a stationary body by a factor of $\gamma$:

$$m = \gamma m_0. \tag{30}$$

In the limiting case, when $\gamma = 0$ ($v = c$), the mass $m = 0$. Therefore, it is probably that the rest mass of photon is equal to zero. In this case, $E_T = E_0$ and $E_R = 0$ correspond to a body being at rest in space. So the rest energy

$$m_0 c^2 = h\nu_{0c} \tag{31}$$

is the internal energy of motion of a body and is caused by any periodic process with a cyclic frequency $\nu_{0c}$ ($h$ is the Plank's constant). In the right part of (31) there is a famous Plank formula for the energy of quantum, and in the left one there also the famous Einstein formula for the rest energy of a body.

When a body moves in space with a velocity $v$, a part of this internal "alive" energy-mass:

$$E_\Delta = E_0 - E_T = (1 - \gamma)E_0 \tag{32}$$

transfers to the motion energy of a body in space. Amounting to the kinetic energy $K$ it forms the second component

$$E_R = E_\Delta + K = E_0 \beta^2 / \gamma = E_0 (1/\gamma - \gamma), \tag{33}$$

shown in Fig. 2 by the curve *5* (we name it as the motion energy of a body in space). In principle, the energy $E_R$ (and together with it the total energy of a body $E$) can increase up to the infinity. But it becomes possible only on account of "pumping" the energy into a body from outside: as it occurs, for example, on accelerators of elementary particles.

## 4. Gravitational Field

### 4.1. Free Fall

Let's consider the processes occurring with a mass $m_0$ of a test body falling freely in the field of other mass $M \gg m_0$. $M$ is considered to be so large that the motion of a test body does not influence on it at all. Let the body begins to fall from infinity, where the gravitational potential (12) is equal to zero. Thereby we consider that our system of two bodies is conservative (absolutely closed), i.e. it is does not exchange by the energy with an environment. Then the total energy of system at any moment is equal to:

$$E = (M + m_0)(c_0)^2 = const. \tag{34}$$

As $M(c_0)^2 = const$, then the energy-mass of a test body does not vary too, i.e. the following condition is satisfied:

$$E_0 = m_0(c_0)^2 = \gamma E_0 + (1 - \gamma)E_0 = E_T + E_\Delta = const . \tag{35}$$

The energy (35) is shown in Fig. 2 by the horizontal line *2*. The line *4* in this Figure shows the change of internal motion energy of a body in time depending on the Lorenz-factor $\gamma$:

$$E_T = \gamma E_0 = (\gamma m_0)(c_0)^2 = m(c_0)^2 . \tag{36}$$

And the motion energy of a test body (fall in the gravitational field):

$$E_\Delta = (1 - \gamma)E_0 = \Delta E_0 = (m_0 - m_0\gamma)(c_0)^2 = \Delta m_0(c_0)^2 \tag{37}$$

is caused by the energy-mass defect

$$\Delta E_0 = \Delta m_0(c_0)^2 , \tag{38}$$

which, according to (14) and (17), is equal to the energy of gravitational coupling:

$$E_\Sigma(R) = m_0|\varphi(R)| = GMm_0/R . \tag{39}$$

In particular, the important consequence follows from the equality $\Delta E_0 = E_\Sigma$:

$$\Delta m_0(R)/m_0 = |\varphi(R)/(c_0)^2| = |\Phi(R)| = 1 - \gamma_{gr}(R). \tag{40}$$

The relations (35) and (40) show that the rest mass $m_0$, which is a kind of the internal energy "accumulator" of a body $E_T$, can transfer to the energy of fall of a test body (37) on the gravitational mass *M*. It will take place at $|\Phi(R)| = 1$ at a distance of gravitational radius

$$R_{gr} = GM/(c_0)^2 , \tag{41}$$

at $v = c_0$. From (37) and (40) it follows

$$\Delta m_0(R)/m_0 = 1 - \gamma_{gr}(R) = 1 - \gamma(R) , \tag{42}$$

that is the equality is fulfilled:

$$\gamma(R) = \gamma_{gr}(R) . \tag{43}$$

### *4.2. Velocity of the Fall*

If we square both parts of (43) and also take into account (8) and (11) then we shall obtain:

$$\beta^2 = 1 - \gamma^2 = 2|\Phi(R)| - (\Phi(R))^2, \qquad (44)$$

or the velocity of free fall itself of a physical body in the gravitational field:

$$v(R) = c_0\sqrt{2|\Phi(R)| - (\Phi(R))^2}. \qquad (45)$$

The formula (45) can be also observed by another way, multiplying both parts of (20) by $(c_0)^2$. In this case we have the relationship:

$$(c_0)^2 = (\gamma c_0)^2 + (\beta c_0)^2 = c^2 + v^2, \qquad (46)$$

from which, with the account of (11), it follows:

$$(v/c_0)^2 = (1 - (c/c_0)^2) = (1 - (1 - |\Phi(R)|)^2) = 2|\Phi(R)| - (\Phi(R))^2. \qquad (47)$$

In the classical mechanics and astronomy the following formula is usually used:

$$v^2 \approx 2|\varphi(R)|, \qquad (48)$$

which describes motion of bodies in weak gravitational fields (for example, in the solar system or in the galaxy) sufficiently well. It automatically follows from the formula (47) as its approximate expression at small velocities of motion of bodies ($v \ll c$).

Fig. 3 plots the dimensionless velocity of fall $\beta$ of a physical body and also its square $\beta^2$ depending on the gravitational potential $|\Phi|$. It is seen, that if the graph for $\gamma(|\Phi|)$ is the inclined line *1* reflecting the linear dependence of the mass of body on the gravitational potential, then the graph $\beta(|\Phi|)$ is the arc of unit circle (curve *2*). The graph $\beta^2$ resembles a sinusoid, which is represented by the curve *3*. The graphs of values $\beta$ (curve *4*) and $\beta^2$ (curve *5*) calculated by the formula (48) are given (dashed curves) for comparison in Fig. 3. These dashed curves reach the unity (velocity of light) at $|\Phi| = 0.5$, and further go away to the region of the superlight velocities, which, according to the relativity theory, can not exist.

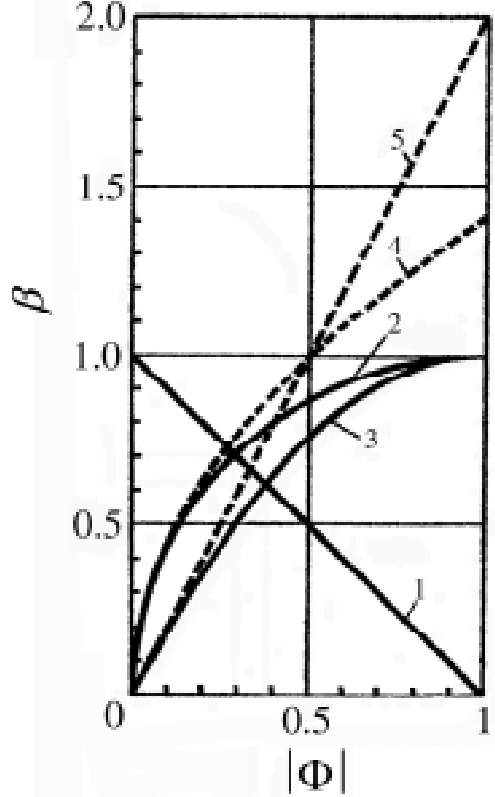

**Fig. 3.** Changes of the velocity of free fall of a body (photon) in the gravitational field depending on its dimensionless potential $|\Phi(R)| = GM/R(c_0)^2$: *1* is the dimensionless velocity of light $\gamma = c/c_0 = 1 - |\Phi(R)|$ (formula 13); *2* is the dimensionless velocity of fall of a body $\beta = v/c_0 = (2|\Phi| - (\Phi)^2)^{1/2}$; *3* is $\beta^2 = 1 - \gamma^2$ (formula 44); *4* and *5* are $\beta = (2|\Phi|)^{1/2}$ and $\beta^2 = 2|\Phi|$, respectively (calculated by the classical formula (48)).

In the equation (45) the velocity of fall of a body can not be more than the velocity of light $c_0$ at any meanings $|\Phi|$. It occurs because that in the right part of (45) there is the second term (negative), which at large $|\Phi|$ limits the increase of velocity $v$ so that it does not exceed $c_0$. As a result, the velocity of fall of a body only tends to the limiting value $c_0$, but can not reach it.

### *4.3. Energy of the Motion*

Let's consider the energy of the motion of test body $E_R$ in more detail. The conservative system of two bodes are of interest for as. From (32), (33) at $K = 0$ (there is no the "pumping" of energy from outside) and (35) we have equalities:

$$E_R(R) = E_\Delta(R) = E_V(R) = E_\Sigma(R) = GMm_0/R , \qquad (49)$$

from which it follows that the energy $E_R$ arises at the expense of energy $E_\Delta$ of the mass defect of this body falling freely from infinity with the kinetic energy $E_V$, equal to the energy of the gravitational coupling $E_\Sigma$ of these bodies. The kinetic energy of the fall in a weak gravitational field (with the velocity $v$ at a distance $R$ from the mass $M$) is found from the formula (48):

$$E_V(R) \approx m_0 v^2(R)/2 \,. \tag{50}$$

The analogous expression for the kinetic energy of the fall of body follows directly from $E_\Delta(R)$:

$$E_V = (1 - \gamma)E_0 = \beta^2 E_0/(1 + \gamma) = m_0 v^2/(1 + \gamma) \,, \tag{51}$$

which at $\gamma \approx 1$ coincides with (50). It indicates that the formulae (32) and (37) quite satisfy to the complementarily principle: at low velocities of motion they transfer to the formula of the classical mechanics for the kinetic energy.

With increase of the velocity of fall the most part of the energy-mass $E_0$ transforms into the energy of motion of the body in space. If the body, for example, with a highest velocity enters the Earth atmosphere and becomes white-hat because of the friction against air, then it will turn into a swift meteorite radiating the bright light. Now, at last, a part of the energy of its motion $E_V$ will radiate as photons into space and will leave our system of bodies. As a result, a meteorite will have the mass defect $\Delta m'$, corresponding to this radiated energy.

As to the part of the energy of motion, which had not time to be radiated, will turn into a thermal energy of the warming-up of meteorite (or the meteorite remnant) after the fall on the Earth surface. Soon this heat will be transferred to the environment. Some time (not so soon) this thermal energy will gradually go into space in the form of the infra-red thermal radiation. Just then the energy (38) will completely leave the system and the mass defect corresponding to it

$$\Delta m_0 = \Delta E_0/(c_0)^2 \tag{52}$$

will finally turn into the energy and radiate from the system.

Let's assume now that at the end of fall from infinity the test body will be captured by the central mass $M$ and will appear on the stationary orbit of radius $R$. It will happen, when the centrifugal ($F_\tau$) and gravitational ($F_G$) forces will be equal to each other:

$$F_\tau = m(v_\tau)^2/R = GMm/R^2 = F_G \,. \tag{53}$$

From here and (49), in particular, follows that kinetic energy of the rotation of a tested body (with the tangential velocity $v_\tau$) is equal:

$$E_\tau = m(v_\tau)^2/2 = E_\Sigma/2 . \tag{54}$$

This expression is known as the virial theorem, which is widely used in the astronomy and other sections of physics. It is considered (see, for example, [5,6]) that the mass in (53) is equal:

$$m = (m_0 \gamma_{gr})/\gamma_\tau = m_0 \gamma' . \tag{55}$$

Here the size (16), describing the reduction of initial mass $m_0$ in the gravitational field appears in numerator. At the expense of the velocity of the orbital motion $\beta_\tau = v_\tau/c_0$ it increases, according to (8), by $1/\gamma_\tau$ раз ($\gamma_\tau = \sqrt{1 - \beta_\tau^2}$). In a final form, with the account of (53), we obtain:

$$\gamma' = \gamma_{gr}/\gamma_\tau = (1 - \beta_\tau^2)/\gamma_\tau = \gamma_\tau . \tag{56}$$

From here it is seen that the energy-mass (55) is less than the initial $E_0$ by the value

$$\Delta E_\tau = (1 - \gamma_\tau)E_0 \approx \beta_\tau^2 E_0/2 = m_0(v_\tau)^2/2 , \tag{57}$$

which is equal to the kinetic energy of rotation of a body (54).

### 4.4. Mass Defect

In description of the motion of body in the gravitational field the classical balance of energy is usually used (see, for example, [4]):

$$E(R) = K(R) + U(R) , \tag{58}$$

where the kinetic energy of motion of a body is

$$K(R) = m_0(v(R))^2/2 \tag{59}$$

and

$$U(R) = - m_0 \varphi(R) \tag{60}$$

is the gravitational energy of field. The negative sign of field is chosen from a condition that the sum (58) at infinity would be equal to zero. Hence it follows that both energies arise as if from nothing because the system is conservative, and at infinity there is neither the field nor motion of a body.

Let's try to understand it in detail in conformity to the gravitational field of the Earth. As a test body we shall use the standard-mass $m_0 = 1$ kg, which we shall place into an orbit of the radius:

$$R_{\tau 1} = 6.376\ 64 \times 10^6 \text{ m} , \tag{61}$$

where the standard rotates with the orbital velocity

$$v_{\tau 1} = 7.907\ 81 \times 10^3 \text{ m/s} . \tag{62}$$

In these calculations the Earth mass $M_\oplus = 5.976 \times 10^{24}$ kg, the gravitational constant $G = 6.672\ 59 \times 10^{-11}$ m$^3$ kg$^{-2}$ s$^{-2}$ and the velocity of light $c = 2.997\ 924\ 58 \times 10^8$ m/s are used [2].

From (49) and (53) the conservation laws follow, in particular:

$$GMm_0 = RE_\Sigma = (v_\tau)^2 R = const . \tag{63}$$

For the Earth and our standard we have:

$$GM_\oplus m_0 = 3.987\ 54 \times 10^{14} \text{ m}^3 \text{ s}^{-2} . \tag{64}$$

The potential energy (60) is shown in Fig. 4 by the curve *1*, which on the orbit of radius $R_{\tau 1}$ (arrow downwards) is equal to:

$$U_1 = 6.253\ 355 \times 10^7 \text{ J} . \tag{65}$$

We have represented it as a positive value, because the energy can never be negative. The kinetic energy (59) of free fall of a body with the velocity (45) at $R = R_{\tau 1}$ has the value:

$$K_1 = 6.253\ 355 \times 10^7 \text{ J}, \tag{66}$$

which coincides with (65), and the velocity of a body itself is equal to:

$$v_1 = 11.183\ 341 \times 10^3 \text{ м c}^{-1} = (\sqrt{2}) v_{\tau 1} . \tag{67}$$

This velocity is known as the second-cosmic velocity, which should be given to a body, in order for it will overcome the Earth gravity and leave the gravitational field of the Earth.

The values (65) and (66) are exactly equal to (37) and (51):

$$\Delta E_{01} = (1 - \gamma_1)E_0 = m_0(v_1)^2/(1 + \sqrt{1 - (v_1/c)^2} = 6.253\ 355 \times 10^7 \text{ J} . \tag{68}$$

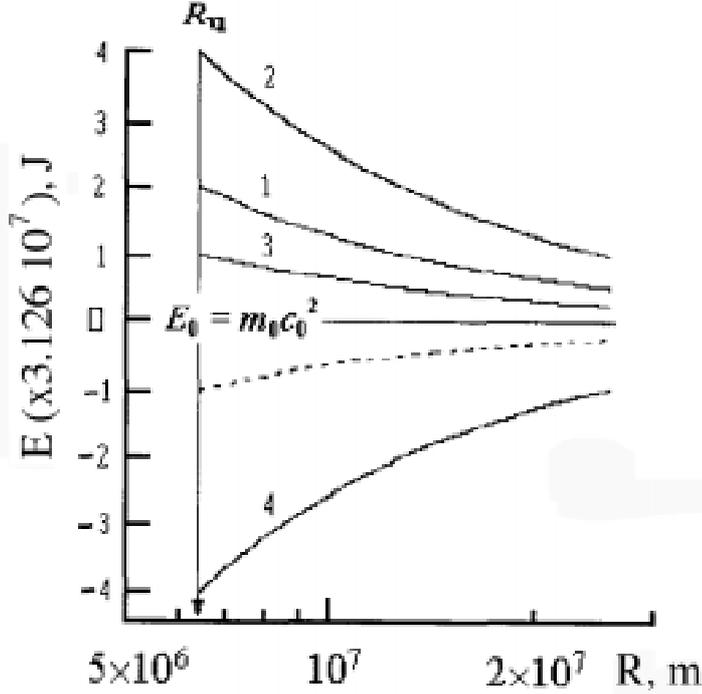

**Fig. 4.** Energy balance of the Earth-standard system (1 kg mass): *1* is the gravitational field energy $|U(R)| = GM_\oplus/R = U_1R_1/R = m_0(v(R))^2/2 = K(R)$; *2* is the energy $E(R) = U(R) + K(R)$; *3* is the kinetic energy of the rotation of 1 kg mass $(K_\tau(R) = m_0(v_\tau(R))^2/2 = K(R)/2)$; *4* is the value $E_0 - E(R) = (1 - \beta^2(R))E_0$ (residual energy-mass of the standard).

It confirms the fact that the mass $m_0$ of the moving body not only can increase by $1/\gamma$, but also decrease during the motion in the conservative system. The formulae (16), (30) and (55) refer to those concrete cases.

It is necessary to notice that the form of writing of the energy balance (58) contains an opportunity of erroneous physical interpretation. The actual equality $U = K$, which contains in it, is usually interpreted as the mutual transition of the potential energy of field into the kinetic energy of motion of the body and back. However, it is one can hardly agree with it.

Let's try to formulate a balance of energy of a freely falling body in the gravitational field more accurately. Let's write the obvious equality:

$$E_0 = (E_0 - m_0\varphi(R)) + m_0\varphi(R) , \qquad (69)$$

coinciding with the equation of gravitational field (14). Here the first term on the right

$$E_0(1 - \varphi(R)/(c_0)^2) = (m_0\gamma_{gr}(R))(c_0)^2 = m(R)(c_0)^2 \qquad (70)$$

is equal to the energy-mass (15) of the stationary body located at a distance $R$ from the central mass $M$, and the second term characterises that positive energy of the gravitational field which we speak about.

When a body falls freely, the kinetic energy (59) arises not from the energy of gravitational field $m_0\varphi(R)$, but at the expense of the additional defect of residual mass $m(R) = m_0\gamma_{gr}(R)$. It is easy to be convinced of it if we add and subtract the kinetic energy (59) in the right part (69):

$$E_0 = (E_0 - m_0\varphi(R) - m_0(v(R))^2/2) + (m_0\varphi(R) + m_0(v(R))^2/2) \,. \tag{71}$$

In this case, if we take into account that (59) and (60) are always equal among themselves, then we shall obtain:

$$E_0 = (E_0 - m_0(v(R))^2) + m_0(v(R))^2 = E_0(1 - \beta^2(R)) + \beta^2(R)E_0 \,. \tag{72}$$

Let's write down this equation in the another form, which directly follows from the equation of motion (20) (for short we omit $R$):

$$E_0 = \gamma^2 E_0 + \beta^2 E_0 = \gamma E_T + \gamma E_R \,. \tag{73}$$

Here the first term (curve *4* in Fig. 4) is equal to:

$$\gamma E_T = m_0\gamma^2(c_0)^2 = (m\gamma)(c_0)^2 = m_0(\gamma_{gr}c_0)^2 = m_0c^2 \tag{74}$$

the energy of rest mass $m_0$ in a given point of the gravitational field with the local light velocity (11), and the second term (curve *2* in Fig. 4)

$$\gamma E_R = \beta^2 E_0 = m_0 v^2 = 2K = 2U \tag{75}$$

reflects the total energy of free fall of the standard in space and the energy of its gravitational coupling with the central mass $M$. It is seen that this energy was formed at the expense of the double mass defect (52):

$$\Delta m = \gamma E_R/(c_0)^2 = (1 - \gamma)(1 + \gamma)m_0 = (1 + \gamma)\Delta m_0 \approx 2\Delta m_0 \,. \tag{76}$$

If we return to our standard ($E_0 = c^2 = 8.987\,551\,787\,368 \times 10^{16}$ J), then at fall into the gravitational field of the Earth from infinity up to the radius (61) it will spend the energy-mass according to (75):

$$\Delta E_1 = \beta_1^2 E_0 = m_0(v_1)^2 = 12.506\,698 \times 10^7 \text{ J} \,, \tag{77}$$

where $v_1$ is the velocity (67). This energy is only $\approx 1.39 \times 10^{-9}$ relative to the total energy- mass of the standard. In the classical mechanics and in a weak gravitational field this mass defect will not practically affect its inert properties.

When the fall will stop, there not all energy (77) but only its half contained in the kinetic energy (66), will leave in the form of radiation. The second its half will remain as before in the system as the energy of the field (39). It is worth noting once again, the energy of the gravitational field and the kinetic energy of the fall of body are of different nature. They do not transform into each other, but are formed and exist independently.

The classical energy balance (58) does not correspond to the real picture even at that time, when it is formed relative to the value $E_0 = m_0(c_0)^2$. In this case it looks like

$$E_0 = E_0 - m\varphi(R) + m(v(R))^2/2 , \tag{78}$$

which does not reveal both the source of the energy of motion and the mechanism of its appearance. The correct answer to this question gives the equation (73). It indicates that the unique source of all kinds of the energy, which has the body in the conservative system, is its initial mass $m_0$. It carries out the role of the distinctive "accumulator" of the giant energy $E_0$, which is used by the nature in order to realize all variety of forms of its existence.

### *4.5. Orbital Rotation of a Body*

Let's find the energy balance of the standard rotating at the stationary Earth's orbit with the velocity (62). If we calculate it from the of the energy-mass defect (57):

$$\Delta E_{\tau 1} = (1 - \gamma_{\tau 1})E_0 = (1 - \sqrt{1 - \beta_\tau^2})E_0 = 3.126\ 672\ 7 \times 10^7 \text{ J} \tag{79}$$

and compare with the kinetic energy of rotation:

$$K_{\tau 1} = m_0(v_{\tau 1})^2/2 = 3.126\ 672\ 8 \times 10^7 \text{ J} \tag{80}$$

then it is possible once again to be convinced of their exact conformity to each other. It speaks that the mass of standard really is a source of energy (80).

However from the energy conservation law follows that the total balance is equal to the sum of values (65) and (80):

$$E_1 = K_{\tau 1} + U_1 = 9.380\ 027 \times 10^7 \text{ J} . \tag{81}$$

It corresponds to the energy-mass defect:

$$\Delta E_1 = (1 - \gamma_{T1})E_0 \approx 3\beta_{\tau 1}^2 E_0/2 = 3K_{\tau 1} \ . \tag{82}$$

The coefficient

$$\gamma_{T1} = \gamma_1 \gamma_{gr1}/\gamma_{\tau 1} = \gamma_1^2/\gamma_{\tau 1} = (1 - \beta_1^2)/\gamma_{\tau 1} \tag{83}$$

is similar to (56), but reflects changes of the mass $m_0$ more correctly. Taking account of the equality $\beta_1^2 = 2\beta_{\tau 1}^2$ in a weak gravitational field it is equal to:

$$\gamma_{T1} = ((1 - \beta_{\tau 1}^2)^2 - \beta_{\tau 1}^4)/\gamma_{\tau 1} = (\gamma_{\tau 1})^3 - \beta_{\tau 1}^4/\gamma_{\tau 1} \approx 1 - 3\beta_{\tau 1}^2/2 \ . \tag{84}$$

In the numerator of (83) there is a value, which reflects a portion of residual mass of the standard near the Earth's surface as a result of its free fall from infinity. Its defect

$$\Delta m_1 = m_0(1 - \gamma_1^2) = m_0 \beta_1^2 \ . \tag{85}$$

corresponds to the energy-mass (77). Note that a half of it leaves the standard irrevocably when the fall in the form of radiation with the energy stops:

$$K_1 = 2E_{\gamma 1} = 2K_{\tau 1} = 6.253\ 355 \times 10^7 \text{ J} \ . \tag{86}$$

In order for the stationary standard after a stopping of fall appears to be on the orbit of radius (61), it should obtain from the outside the kinetic energy (80) equal to a half of the lost energy (86). This energy will rotate the standard with the first space velocity (62), and in this case it will return to the mass $m_0\gamma\gamma_{gr}$ again and will increase it by $1/\gamma_{\tau 1}$ according to (27). Generally the total energy, which is released from the mass of standard in the form of the kinetic energy of its rotation in the closed orbit of radius $R$ and the gravitational interaction, is equal to:

$$\Delta E_\tau(R) = m\varphi(R) + m(v_\tau(R))^2/2 = 3E_\tau(R) \ .$$

Besides, the energy

$$E_\gamma(R) = E_\tau(R) = m(v_\tau(R))^2/2 \tag{87}$$

will still remain lost for the conservative system (curve *3* in Fig. 4). It forms the mass deficit (dashed curve in Fig. 4) which will return to a body only in that case, when it obtains one more portion of the energy $E_\tau$ from the outside additionally, in order for it will go away from the gravitational mass *M* to infinity. In this case the body itself will have always the energy in the form of

$$E(R) = E_0 - 4E_\tau(R),  \tag{88}$$

shown in Fig. 4 by the curve *4*.

## 5. Atom of Hydrogen

The change of mass of a body in conservative macrocosm has been discussed above. The similar picture is observed and in microcosm. Consider it on the example of the atom of hydrogen which is the most abundant element of the Universe. For the description of its basic properties the Rutherford-Bohr atom model is used [7,8], in which the electron moves with kinetic energy

$$K(R) = m_e(v_\tau(R))^2/2 \tag{89}$$

in the Coulomb field

$$\varphi(R) = - e/4\pi\varepsilon_0 R = - ke/R \tag{90}$$

of the stationary point proton, where $\varepsilon_0 = 8.854\ 187\times10^{-12}$ F m$^{-1}$ is the electric constant; $k = 8.987\ 551\times10^9$ m F$^{-1}$. The negative sign in (90) is chosen from a condition, in order for the total energy

$$E(R) = K(R) + e\varphi(R) = m_e(v_\tau(R))^2/2 - ke^2/R \tag{91}$$

at the indefinitely long distance *R* between the electron and proton would be equal to zero. The balance of energy (91) is similar to (58). In this similarity the actual uniformity of both systems is reflected. There is an opinion (see, for example, [5,9,10]) that the structural ladder of matter (from infinitesimal up to indefinitely large) is subordinated to a "matryoshky" principle, i.e. it is constructed by inserting of small structures into larger ones.

Bohr assumed the existence in atoms of the certain stationary conditions (stationary orbits), in which electrons do not radiate (1-st postulate). According to his second postulate electrons can be only at such "allowed" orbits, for which the multiplicity conditions of the angular momentum of an electron to the Plank's constant *h* are carried out:

$$2\pi R_n m_e v_n = nh, \tag{92}$$

where $n$ is the principal quantum number ($n = 1, 2, 3, ….$). From the classical and quantum mechanics under condition of (92) follows that the electron energy and the radii of "allowed" orbits can be only of discrete values. In this case the numbers $n$ are clearly interpreted as the numbers of orbits. Radius of $n$-th electron orbit in the hydrogen atom appears to be

$$R_{en} = R_{e1}n^2, \qquad (93)$$

where

$$R_{e1} = h^2/4\pi^2 ke^2 m_e = 5.291\ 772\ 552\ 747\times 10^{-11}\ \text{m}, \qquad (94)$$

and its energy of rotation around of the proton is:

$$K_{en} = -E_{e1}/2n^2, \qquad (95)$$

where

$$E_{e1} = (2\pi ke^2/h)^2 m_e = 27.211\ 394\ \text{eV}. \qquad (96)$$

Bohr supposed further (3-rd postulate), that the radiation occurs only in that case, when the electron "jumps" from one stationary orbit to another with smaller energy. In this case the photon is radiated with the energy

$$h\nu_{nm} = K_{en} - K_{em} = (E_{e1}/2)(1/m^2 - 1/n^2). \qquad (97)$$

Usually it is considered (see, for example, [7,8]) that the energy $E_{e1}/2 = 13.605\ 697$ eV corresponds to the depth of potential well, in which the electron is in the nonexited state ($n = 1$). Formally the energy of well (91) is negative. It means, that it has left the electron-proton system in the hydrogen atom. So much energy ($E_{\gamma 1} = E_{e1}/2$) is radiated as light in the moment of formation of the atom from the proton and electron (recombination radiation) [1,8]. As a result, the mass of hydrogen atom appears less then the initial sum of masses of proton and electron by a magnitude equal to the mass defect:

$$\Delta m_{e1} = E_{\gamma 1}/c^2. \qquad (98)$$

When the hydrogen atom is ionized (to tear it to a proton and electron), then the atom obtains from outside the same quantity of energy named in this case as ionization energy of the atom [7,8].

Further we use not the balance of energy (91), which as well as (58) does not correspond to a real picture, but the complete balance:

$$E_e = m_ec^2 = (E_e - \Delta E_e) + \Delta E_e = const, \quad (99)$$

which is equal to the total energy of the stationary electron removed from a proton to infinity. Here the value

$$\Delta E_e(R_{en}) = E_\gamma(R_{en}) + K(R_{en}) + U(R_{en}) = E_\gamma(R_{en}) + m_e(v_\tau(R_{en}))^2/2 + ke^2/R_{en} \quad (100)$$

reflects the total energy-mass, which is released from the mass of electron for the radiation $E_\gamma(R_{en})$ (leaves the atom of hydrogen), the kinetic energy $K(R_{en})$ of rotation of the electron around the proton in an orbit of radius $R_{en}$ and for the energy of Coulomb interaction $U(R_{en})$. The value in brackets in (99) is equal to the energy-mass which the electron has in the form of its residual mass. It is seen from (92), that in his model of the atoms Bohr actually postulated the existence of elementary action (angular torque)

$$D_{e1} = 2\pi R_{e1} m_e v_{e1} = h = 4.135\ 669\ 246\ 498 \times 10^{-15}\ \text{eV s}, \quad (101)$$

which is accomplished by the World medium to the electron for its one rotation around the hydrogen atom nucleus in the nonexited state ($n = 1$) with the velocity $v_{e1}$ at the length of a circle $2\pi R_{e1}$. The electron makes $\nu_{e1}$ actions per 1 s:

$$\nu_{e1} = 1/T_{e1} = v_{e1}/2\pi R_{e1} = 6.579\ 683\ 786\ 153 \times 10^{15}\ \text{s}^{-1} \quad (102)$$

and has the energy of motion:

$$E_{e1} = \nu_{e1} D_{e1} = \nu_{e1} h = 27.211\ 395\ \text{eV}. \quad (103)$$

Hence it follows another definition of action (101): that is an elementary work, which is made by the World medium to the electron for its one rotation around the hydrogen atom nucleus in the nonexited state. In this case, the electron moves along a circular orbit of radius $R_{e1}$ with the velocity

$$v_{e1} = 2.187\ 691\ 402\ 410 \times 10^6\ \text{m s}^{-1}. \quad (104)$$

Dimensionless value of this velocity

$$\beta_{e1} = v_{e1}/c = \alpha = 7.297\ 353\ 032\ 177 \times 10^{-3} \approx 1/137 \quad (105)$$

is the constant $\alpha$ of fine-structure. If we multiply the rest energy-mass of an electron by $\alpha^2$, then we obtain:

$$\Delta E_{e1} = (m_e c^2)\alpha^2 = m_{e1}(v_{e1})^2 = 2K(R_{e1}) = \Delta m_{e1} c^2 = 27.211\ 397\ \text{eV} \quad (106)$$

in complete conformity with (96) and (103) up to the sixth sign after a comma. It indicates that the sum of energies $E_\gamma(R_{e1}) + K(R_{e1})$ in balance (100) was really formed at the expense of the mass defect of electron at its motion in space with the velocity (104).

The Coulomb energy of attraction of the electron and proton at a distance $R_{e1}$ is equal to:

$$U_{e1} = k(e^2/R_{e1}) = 27.211\ 394\ \text{eV}\ , \tag{107}$$

It coincides with the energy (103) and the mass defect (106). The value (107) is called as the binding energy of the electron and proton in the nonexited condition of hydrogen atom and it is usually correspond to the double kinetic energy of the electron rotation (95)

$$U(R_{en}) = 2K(R_{en})\ , \tag{108}$$

considering it as the manifestation of virial theorem (54) in the Coulomb field [8]. In view of the above mentioned we can now express the energy (100) more generally:

$$\Delta E_e(R_{en}) = 2U(R_{en}) = 4K(R_{en}) = 2m_e(v_{en})^2 = 2m_e(v_{e1}/n)^2 = 2E_{e1}/n^2\ , \tag{109}$$

where the velocity of orbital rotation of the electron $v_{en}$ is determined by the formula [7,8]:

$$v_{en} = 2\pi k e^2/hn = v_{e1}/n\ . \tag{110}$$

The energy (109) corresponds to the total mass defect:

$$\Delta m_e(R_{en}) = \Delta E_e(R_{en})/c^2 = 2m_e(\beta_{en})^2 = 2m_e(\alpha/n)^2\ . \tag{111}$$

In the ground state the relative electron mass defect is equal to of $\Delta m_{e1}/m_e = 2\alpha^2 \approx 10^{-4}$. With the growth of number of the orbit it quickly decreases and tends to zero at ionization of the hydrogen atom.

## 6. Discussion

The conservative systems considered by us especially different on their spatial scale and nature of working forces have common features. Especially it is clearly seen in the Table, where the basic formulae and components of balance of energy of both systems are presented. Between them there is an almost complete analogy, except quantum properties of the electron, manifested in units of the Planck constant. However here again the positions are drawn together. It is expressed in a manifestation of quantum properties of gravitation (see, for example, [5]).

**Table**. Comparative characteristics of energy balance of conservative systems consisting of two bodies in macro- and microcosm.

| Standard (1 kg) on the Earth orbit (the first space velocity) | | Atom of hydrogen (ground state) | |
|---|---|---|---|
| Kind of energy | Value of energy (Joule) | Kind of energy | Value of energy (eV) |
| $U_1 = GMm_0/R_{\tau 1}$ | $6.253\,355 \times 10^7$ | $U_{e1} = k(e^2/R_{e1})$ | 27.211 394 |
| $K_{\tau 1} = m_0(v_{\tau 1})^2/2$ | $3.126\,672 \times 10^7$ | $K_{e1} = m_{e1}(v_{e1})^2/2$ | 13.605 697 |
| $E_{\gamma 1} = K_{\tau 1}$ | $3.126\,672 \times 10^7$ | $E_{\gamma 1} = K_{e1}$ | 13.605 697 |
| $K_1 = m_0(v_1)^2/2$ | $6.253\,355 \times 10^7$ | $E_{e1} = m_{e1}(v_{e1})^2 = \nu_{e1}h$ | 27.211 395 |
| $2\Delta E_{\tau 1} = \beta_{\tau 1}^2 E_0$ | $6.253\,344 \times 10^7$ | $\Delta E_{e1} = \alpha^2 E_e$ | 27.211 397 |
| $\Delta E_1 = 2\beta_{\tau 1}^2 E_0 = 4K_{\tau 1}$ | $12.506\,698 \times 10^7$ | $\Delta E_1 = 2\alpha^2 E_e = 4K_{e1}$ | 54.422 788 |

Let's add some common reasons to the above. As any body consists of atoms and molecules, and those, in turn, of elementary particles (electrons, protons etc.), then it is naturally the question is about the mass of these particles, because their quantity does not change when the body moves.

Let's return to Fig. 3 and we shall look the relation of the dimensionless velocity of light (line 1) to dimensionless velocity $\beta$ of fall of a body (among then photon) in the gravitational field found by the formulas (11) and (44). It is seen that when the dimensionless gravitational potential $|\Phi(R)|$ is equal to $1 - 1/\sqrt{2}$, the velocity of fall $v$ reaches the value of local velocity of light with $c = c_0/\sqrt{2}$. It occurs at a distance of $3.41R_{gr}$ ($R_{gr}$ is the gravitational radius (41) of a "black hole") from gravitating centre (point of intersection of curves 1 and 2). At distances to the "black hole", shorter than mentioned one, the velocity of fall of a body becomes more than the local velocity of light. A body and photon as if change by roles, when they fall into the nearest vicinities of "black hole". In this case, the photon becomes heavy and inactive as an usual body, and the body (more correctly those elementary particles, of which it consists) – easy and fast as the photons.

Let's look at (23) once again, which characterizes not only certain geometric properties of four-dimensional space-time, but also, probably, has a deeper and wider physical sense. If we divide the both parts of (23) by $(\Delta t)^2$ and take into account $\Delta t_0/\Delta t = \gamma$, then we obtain the equation:

$$c^2 = (\gamma c)^2 + v^2, \qquad (112)$$

which, after the multiplication of both parts by $m = m_0/\gamma$, becomes (28):

$$E = E_0/\gamma = \gamma m_0 c^2 + mv^2 = E_0\gamma + E_0\beta^2/\gamma = E_T + E_R . \qquad (113)$$

We have already discussed above the sense of this equation and of both components $E_T$ and $E_R$ (see section 4). It emphasizes once again the intimate relationship between the energy of internal motion of a body $E_T$ and its own course of the time $\Delta t_0$.

Let's find the relativistic kinetic energy of motion of a body of mass $m_0$ and velocity $v$. It follows from the equation (25) as a difference between the total energy of moving body $E$ and its energy of rest $E_0$:

$$K = E - E_0 = E_0(1 - \gamma)/\gamma = E_0\beta^2/\gamma(1 + \gamma) = E_R/(1 + \gamma) . \qquad (114)$$

From here, in particular, at $\gamma \approx 1$ we obtain its the classical expression:

$$K = m_0 v^2/\gamma(1 + \gamma) = mv^2/(1 + \gamma) \approx m_0 v^2/2 . \qquad (115)$$

If we substitute $E_0$ from (31) in (113)

$$E = h\nu_{0c}/\gamma = h\nu_c = h(\gamma\nu_{0c}) + mv^2 , \qquad (116)$$

then one can see that the energy of motion of a body in space

$$E_R = mv^2 = h\nu_{0c}\beta^2/\gamma = h\nu_c\beta^2 = h\nu_v , \qquad (117)$$

is also of quantum nature caused by the frequency

$$\nu_v = (c/\lambda_c)\beta^2 = v^2/c\lambda_c = v/\lambda_v , \qquad (118)$$

which reflects the additional increase of $E_0$ on account of the kinetic energy (115)

$$K = E_R/(1 + \gamma) = h\nu_v/(1 + \gamma) . \qquad (119)$$

The ratio of the frequency (118) to the Compton frequency $\nu_{0c} = E_0/h$

$$\nu_v/\nu_{0c} = \beta^2/\gamma = 1/\gamma - \gamma \qquad (120)$$

is very small at $\gamma \approx 1$ and reaches the greatest values in the case of relativistic velocities (at $\gamma \approx 0$). The formula (120) refers to the nonclosed system and follows directly from (33). To find the similar ratio for the conservative system (for example, for the gravitational field), one can use the equation (35):

$$E_0 = \gamma E_0 + mv^2/(1 + \gamma) , \qquad (121)$$

from which and the equation (44) the dependence follows:

$$\nu_v/\nu_{0c} = \beta^2 = 2|\Phi(R)| - (\Phi(R))^2 , \qquad (122)$$

shown in Fig. 3 by the curve 2.

De Brogue guessed about the quantum nature of energy (117) in his time. He put forward a hypothesis that the corpuscle-wave duality is restricted to all to elementary particles without exception (and even macrobodies) in any forms of their motion. Actually he expanded the concept of elementary action (101) for all cases of the motion:

$$D = h = 2\pi R_v mv = \lambda_v p , \qquad (123)$$

which is made by the World medium over a body of mass $m$ per one period $T_v = 1/\nu_v = \lambda_v/v$ of the off-beat cyclic field-mass transformation of matter; $p = mv$ is the momentum of a body. The generalized action (123), as applied to an electron, results in the following equalities:

$$R_{ec}m_e c = R_{e1}m_e v_{e1} = R_{ev}m_e v_e = h/2\pi , \qquad (124)$$

in which the absolutely identical quantum character of the motion of matter in itself electron, its energy of rotation in the hydrogen atom (in the ground state) and in the free motion outside of the atom is reflected. If we rewrite the left part of (124) as

$$m_e c = h/2\pi R_{ec} \qquad (125)$$

and multiply the both his parts by the velocity of light $c$, then we obtain the equation (31) for the total energy of rest electron:

$$m_e c^2 = h\nu_{ec} = E_e = 5.109\,990\,527 \times 10^5 \text{ eV} , \qquad (125)$$

where $c/2\pi R_{ec} = \nu_{ec} = 1.235\,589\,842 \times 10^{20}$ s$^{-1}$. The length of circle $2\pi R_{ec} = \lambda_{ec}$ is named as the Compton wavelength of electron:

$$\lambda_{ec} = h/m_e c = 2.426\,310\,596 \times 10^{-12} \text{ m} \approx 0.0243 \text{ A} . \qquad (126)$$

For the first time de Brogue obtain the relationship of both parts of the famous equation (125) as far back as 1920s of the last century. He already in that time tried to find a common beginning between the total energy of particle and the mysterious frequency of internal processes in it, leading to the occurrence of corpuscle-wave duality [11].

Conformably to all kinds of the motion of an electron we have:

$$E_e/\nu_{ec} = E_{e1}/\nu_{e1} = E_{ev}/\nu_{ev} = h . \qquad (127)$$

The Heisenberg uncertainty principle follows from (124) and (127):

$$pR = (p\eta)(R/\eta) = \Delta p \Delta R = h/2\pi , \qquad (128)$$

$$Et = (E\eta)(t/\eta) = \Delta E \Delta t = h/2\pi , \qquad (129)$$

where $t = R/c$; $\eta$ is the arbitrary value. It is based on the assumption of indivisibility of the elementary action (123). The physical sense of this action and the mechanism of its realization remain yet unknown up to the end. They are probably connected with one of the most fundamental properties of matter going deeply into its structure up to the Planck length [12]:

$$l_h = \sqrt{hG/c^3} \approx 4.05 \times 10^{-35} \text{ m} . \qquad (130)$$

The nature of the mass is one of major unsolved task of the modern physics [13]. It is accepted to consider that the mass of an elementary particle is determined by the fields, which are connected to it (electromagnetic, nuclear etc.), however, nobody yet managed to create the quantitative theory of the mass [13]. Therefore, any information about new properties of the mass throwing the additional light on this problem are important here.

## 7. Conclusion

We have considered changes of the mass within the framework of the relativity theory a bit of wider, than it is usually done. Our research is based on the Lorenz transformation (8) and the Einstein equation (11). The equations of the motion (20) and (13) obtained on their basis have appeared to be rather effective in the solution of a number of the difficult tasks connected with the dynamics of the interacting bodies in the conservative system. The equation (20) has allowed to open in the formula (28) two kinds of the energy of body (see Fig. 2): the energy of internal motion in time (29) and the energy of motion of a body in space (33).

The role of the mass has appeared to be more important, than it is determined by the formula (27). This formula directly follows from the relativistic equation (4). It indicates that the mass of a moving body increases as its energy increases. The fact of the growth of the energy-mass has been reliably established for the elementary particles in the numerous experiments on the accelerators and in cosmic rays. However, we pay attention to the motion of a body in the absolutely closed system, where the energy does not arrive from the outside. Here the mass manifests itself not only as a measure of the inert and gravitational properties of a body, but also as the source of energy of the motion of body (see section 4).

Apparently, those researchers mistake who believe, that the total energy of body at its fall in the gravitational field increases by the relativistic law (4). In fact, according to the graviton model, explaining the gravitational attraction as the exchange by hypothetical gravitons [1,5,8,9], the total energy-mass of body in the conservative system should remain constant. Under such conditions a single source of energies of the gravitational interaction and the motion of a body can be only its mass of rest $m_0$. It plays a role of the peculiar "accumulator" of giant energy $E_0 = m_0 c^2$, which is used by the nature to realize all variety of forms of existing.

The similar picture takes place in the conservative system, apparently, at different levels of the structural ladder of the matter. We have tried to show it by concrete examples of the fall of the 1 kg standard in the gravitational field of the Earth and the formation of the hydrogen atom from an electron and a proton at their approach from infinity. This similarity is very clearly seen in the Table, where the basic formulae and components of the energy balance of both systems are given. Between them there is an almost complete analogy.

Let's address once again to the spatial-time interval (24), which is usually interpreted as the geometric factor of the curved four-dimensional space-time [1]. From our point of view, it can be of another physical sense (Fig. 1): it is some "distance in time", which a body passes for the time $\Delta t$, expressed due to the coefficient $c$ in same units of length, as the distance $\Delta r = v\Delta t$, covered by this body in space for the same time $\Delta t$, counted by the clock of an observer, relative to which the body moves. The basis for (24) is the Pythagor theorem connecting the distances of four-dimensional space-time in the equation (23). It is not excluded that the interval (24) is connected not only with the geometry of space-time, but also with the energy of motion of matter. The sense of the last assumption is opened by the equations (112) and (113).

## Acknowledgments

This work was supported by the Russian Foundation for Basic Research, project no. 05-02-17857a.